\documentclass[pra,twocolumn]{revtex4-1}
\usepackage{float,graphicx, color, dsfont}
\usepackage{amsmath}
\usepackage{amssymb}
\usepackage{makeidx}
\usepackage{amsfonts}
\usepackage{graphics}
\usepackage[ansinew]{inputenc}
\usepackage{comment}

\usepackage{color}
\usepackage[colorlinks,hyperindex]{hyperref}

\usepackage{mathrsfs}
\hypersetup
{
	colorlinks,%
	citecolor=black,%
	linkcolor=black,%
	urlcolor=black,%
}

\newcommand{\ket}[1]{\left| #1 \right\rangle}
\newcommand{\bra}[1]{\left\langle #1 \right|}

\begin{document}
\title{Non-Gaussian operations in measurement device independent quantum key distribution}
\author{Jaskaran Singh}
\email{jsinghiiser@gmail.com}
\affiliation{Department of Physical Sciences, Indian
Institute of Science Education and Research (IISER)
Mohali, Sector 81 SAS Nagar, Manauli PO 140306 Punjab India.}
\author{Soumyakanti Bose}
\email{soumyakanti.bose09@gmail.com}
\affiliation{Department of Physical Sciences, Indian
Institute of Science Education and Research (IISER)
Mohali, Sector 81 SAS Nagar, Manauli PO 140306 Punjab India.}

\begin{abstract}
Non-Gaussian operations in continous variable (CV) quantum key distribution (QKD) have been 
limited to photon subtraction on squeezed vacuum states only. This is mainly due to the 
ease of calculating the covariance matrix representation of such states. In this paper we 
study the effects of general non-Gaussian operations corresponding to photon addition,
catalysis and subtraction on squeezed coherent states on CV measurement device 
independent (MDI) QKD. We find that non-Gaussianity coupled with coherence can yield significantly 
longer transmission distances than without. Particularly we observe that zero photon catalysis on 
two mode squeezed coherent state (TMSC) is an optimial choice for CV MDI QKD, while single 
photon subtraction is also a good candidate; both of them offering nearly $70$ km
of transmission distances. We also derive a single generalized covariance matrix for the aforementioned
states which will be useful in several other aspects of CV quantum information processing. 

\end{abstract}

\maketitle

\section{Introduction}

Quantum key distribution (QKD)~\cite{quantum_crypt,arxiv_qkd_review} is one of the most widely known and commercially available 
application of quantum information theory. 
It provides a measure of security that is not possible 
to achieve using classical key distribution schemes. While the latter protocols are traditionally deemed 
secure by virtue of some computationally hard to solve mathematical problem, the security 
of former is based on a principle of nature like 
Heisenberg uncertainity principle~\cite{bb84,shor_sec_proof_bb84}, 
no cloning theorem~\cite{ralph_cvqkd,hillery_cvqkd,cerf_cvqkd,grosshans_cvqkd} and 
Bell's theorem~\cite{ekert_e91,acin_bell_qkd,pawloski_bell_qkd}. Ideally, QKD protocols are 
unconditionally 
secure~\cite{renner_sec_proof_cvqkd,anthony_sec_proof_cvqkd,
	anthony_sec_proof_cvqkd2, anthony_sec_proof_cvqkd3}, but 
noise in the measurement and preparation devices may cause the security to be entirely compromised. For this
purpose certain assumptions/pre-conditions have to be imposed on all the devices available to the parties, Alice and Bob.
However, in order for the protocol to be practical, it is desirable for the assumptions to be minimal. 
For example, the standard BB84 protocol assumes that the parties share a single qubit state and 
have access to dichotomic measurements only.

Among all, measurement device independent (MDI) is a prominent class of QKD protocols~\cite{lo_dvmdiqkd,braunstein_dvmdiqkd,curty_dvmdiqkd_theory,ottaviani_dvmdiqkd_theory,
	dcmdiqkd_exp,dvmdiqkd_exp2, dvmdiqkd_exp3} , based on entanglement swapping, that work under 
the assumption that the state 
preparation devices with Alice and Bob are well characterized such that an eavesdropper, Eve has no 
access to any side channels, while the measurement devices are uncharacterized and untrusted. With 
this, continous variable (CV) MDI QKD protocols~~\cite{li_cvmdiqkd,ma_cvmdiqkd,pirandola_cvmdiqkd_exp, zhang_cvmdi_qkd, papanastasiou_cvmdiqkd_theory,
	lupo_cvmdiqkd_theory,pirandola_cvmdiqkd_exp} further boasts of longer transmission
distances that can encompass a small metropolitian city, in comparison to the discrete variable counterparts~\cite{gehring_cvmdi_qkd_exp,xu_cvmdi_qkd}.

While CV MDI QKD has been well studied using Gaussian states like two mode squeezed vacuum (TMSV), 
few recent studies have shown that non-Gaussianity \cite{zhang_photon_sub_entanglement,zhao_cvmdiqkd_virtual_photon_sub,cvcatalysis,ma_cvmdiqkd_photon_sub,
	subtraction13,virtualpra16,ye_cvmdi_qkd_catalysis,wang_cvmdi_qkd_catalysis,hu_cv_qkd_nongaussian} and coherence \cite{chandan_cvmdiqkd} can have a major impact on maximizing the transmission 
distances. 
It might be noted that the desired non-gaussianity could be induced in many ways such as photon subtraction, photon addition, catalysis {\em etc.}
While the case of photon subtraction has been studied in full depth, the process of photon catalysis has been explored to some extent. 
Furthermore, the impact of coherence on CV MDI QKD protocols has only been made possible in the case of photon subtraction,
where it was shown to be quite advantageous~\cite{chandan_cvmdiqkd}. 
However, a general treatment of these non-Gaussian processes along with fiducial coherence has been not been attempted due to the difficulty in obtaining closed form solution for the covariance matrix - a primary ingredient in Gaussian modulated CV QKD.

In this paper we derive a generalized covariance matrix for processes corresponding to photon addition,
photon subtraction and catalysis on two mode squeezed coherent states 
(PATMSC, PSTMSC and CTMSC respectively), which to the best of our knowledge has not 
been attempted before. The covariance matrix takes into account the number of 
photons added, subtracted or catalysed as parameters which can be chosen 
arbitrarily. Furthermore, we introduce displacement (coherence) as a parameter too, which can 
also be chosen arbitrarily. Coupled with all the parameters, our covariance 
matrix is the most general one attempted till date and apart from its immediate application in 
QKD, it is expected to be of immense 
interest in other non-Gaussian information processing tasks such as quantum teleportation, entanglement swapping~\cite{marshall_cv_entanglement,fabio_cv_nongaussian_entanglement}, quantum internet~\cite{su_chinese} {\em etc} .

We provide a detailed discourse on the impact of non-Gaussianity coupled 
with small displacements on CV MDI QKD protocols. Such a discourse would be of 
immense interest to experimentalists in selecting the most optimal state for QKD based on transmission
distances, noise robustness and/or coherence. While it has already been shown that a small 
amount of coherence is a actually helpful process in the case of 
photon subtraction~\cite{chandan_cvmdiqkd}, we find that it 
is true for photon addition and catalysis too. This reinforces the idea that non-Gaussianity coupled with 
coherence leads to better performance in CV MDI QKD. Particularly, we find that CTMSC state outperforms the other states 
in both the aforemetioned criterion allowing transmission distances of almost $70$ Km, while 
photon subtraction is equivalently a good candidate too. We also find that addition of photons 
is not an adequate process to introduce non-Gaussianity in CV MDI QKD protocols as it does not 
lead to any significant increase in transmission distances.

The paper is organized as follows: 
In Sec.~\ref{sec:schematic_ngqkd} we provide a brief idea of CV MDI QKD with Gaussian states followed by corresponding cases with non-gaussian operations such as photon subtraction, addition and catalysis. 
In this section we also discuss the process to calculate secure key rates.
Sec.~\ref{sec:results} presents our simulation results on experimentally obtainable keyrates.
In Sec.~\ref{sec:conclusion} we summarize our results.

\begin{figure*}
	\includegraphics[scale=1]{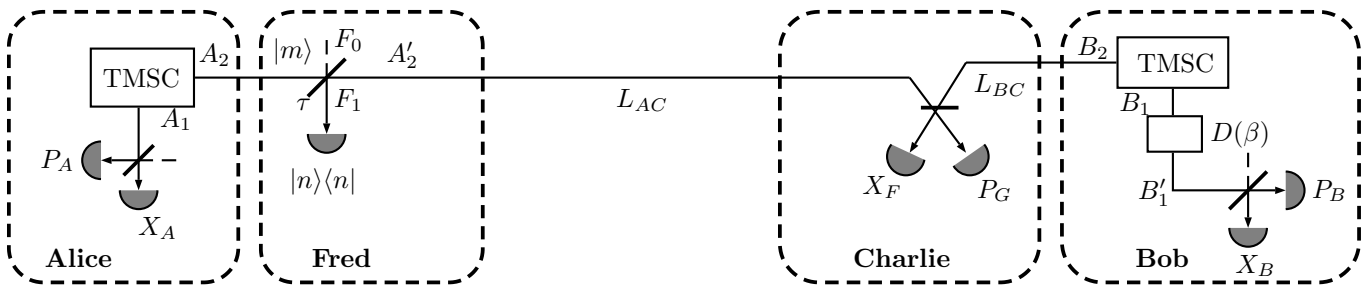}
	\caption{Scheme to implement CV MDI QKD using non-Gaussian states. Two trusted 
		parties Alice and Bob 
		produce TMSC states while a third untrusted party Fred performs photon addition, subtraction or 
		catalysis. A fourth untrusted party Charlie performs homodyne measurements on the two modes
		obtained from Fred and Bob after mixing them via a BS. The results of Charlie's measurements are 
		declared publicly. }
	\label{fig:qkd}
\end{figure*}

\section{CV MDI QKD using CTMSC, PATMSC and PSTMSC}
\label{sec:schematic_ngqkd}

In this section we first provide a brief overview of CV MDI QKD using Gaussian states. 
We then describe the scenario where non-Gaussian states can be utilized. We then elucidate 
how secure keyrates are to be computed.

\subsection{Gaussian CV MDI QKD}
\label{subsec:gs_cv_mdi_qkd}

Consider two parties Alice and Bob who wish to share a secure key. Each party prepares a TMSV state
with quadrature variances $V_A = V_B$ respectively. The two modes with each party are labelled as 
$A_1$, $A_2$ and $B_1$, $B_2$ respectively. Alice and Bob transmit the modes $A_2$ and $B_2$ to 
a third untrusted party, Charlie, while retaining the modes $A_1$ and $B_1$ with themselves. These
modes are transmitted via quantum channels of length $L_{AC}$ and $L_{BC}$ respectively. The total 
transmission distance between Alice and Bob is then $L=L_{AC}+L_{BC}$. 

Charlie interferes the two modes with the help of a $50:50$ beam splitter (BS) which has two 
output modes $C$ and $D$. He then performs a homodyne measurement of $x$ quadrature on $C$ and 
$p$ quadrature on $D$ to obtain outcomes $X_C$ and $P_D$ respectively. The obtained outcomes 
$\lbrace X_C,P_D\rbrace$ are then publicly announced by Charlie. Subsequently, Bob performs 
a displacement operation $D(\alpha)$ on his 
retained mode $B_1$ to get $B'_1$, where $\alpha = g(X_C+iP_D)$ and $g$ is the gain factor.

After these operations, the modes $A_1$ and $B'_1$ are found to be entangled. Alice and Bob 
then perform heterodyne measurements on their entangled modes to obtain the outcomes
$\lbrace X_A, P_A\rbrace$ and $\lbrace X_B,P_B\rbrace$ which are correlated. The scheme is 
given in Fig.~\ref{fig:qkd} with the exception of Fred.

Finally, both the parties perform information reconciliation and privacy amplification
to obtain a secure key.

\subsection{CV MDI QKD using non-Gaussian states}
The scenario of CV MDI QKD which utilizes non-Gaussian states is quite similar to the Gaussian CV MDI QKD, with
the exception of an additional untrusted party Fred who acts on the mode $A_2$ as shown in 
Fig.~\ref{fig:qkd}. We also assume that Bob performs reverse reconciliation (RR), which
implies that his outcomes are taken to be as a reference for Alice to reconcile with.

We describe the basic scheme of our protocol with relevant calculations done in the Appendix. 
We make use of phase space methods (particularly Wigner functions) to perform the calculations. The
protocol proceeds as follows:

\textbf{Step 1:} Alice prepares a TMSC state $|\psi\rangle_{A_1 A_2 }$ with 
quadrature variance $V_A = \cosh(2 r)$. Such a state can be achieved by using a non-linear optical 
downconverter and the process is described as
\begin{equation}
|\psi\rangle_{A_1 A_2 } = S_{12}(r)D_1(d)D_2(d)|00\rangle,
\label{eq:psi}
\end{equation}
where $S_{12}(r) = \text{exp}[r(\hat{a}^\dagger_{A_1}\hat{a}^\dagger_{A_2}-\hat{a}_{A_1}\hat{a}_{A_2})]$
is the squeezing operator with parameter $r$ while $D_i(d) =
\text{exp}[d ( \hat{a}^\dagger_{A_i} - \hat{a}_{A_i} ) ]$ 
is the displacement operator displacing mode $A_i$ only
along the $x$ quadrature with magnitude $d$.

\textbf{Step 2:} Alice then transmits the mode $A_2$ to the untrusted party
Fred,  who mixes it with the mode $F_0$ through a BS with transmittivity $\tau$. 
The mode 
$F_0$ is initialized in the state $|m\rangle\langle m|$. The corresponding transformation $\mathcal{U}_{A_2F_0}^{BS}$ is 
described as 
\begin{equation}
\mathcal{U}_{A_2F_0}^{BS}: |\psi\rangle_{A_1A_2}|m\rangle_{F_0} \rightarrow 
|\Psi\rangle_{A_1A'_2F_1}.
\end{equation}

Using a photon number resolving detector (PNRD), Fred then performs a projective measurement
$\lbrace |n\rangle\langle n|, \mathds{1}-|n\rangle\langle n|\rbrace$ on the mode $F_1$, where
$|n\rangle\langle n|$ corresponds to $n$ photons being detected.  
As a consequence,  for the modes $A_1$ and $A'_2$,  we  call the resultant two mode state as $(m,n)$-TMSC and is given by the unnormalized state
$|\Psi\rangle_{A_1A'_2}^{(m,n)} = 
{}_{F_1}\langle n|\Psi\rangle_{A_1A'_2F_1}$.  
The normalization is the probability of $n$ photon detections and is given as
\begin{equation}
P^{(m,n)} = \sum_{r}\sum_{s}|{}_{A_{1}}\langle r|
{}_{A_{2}^{'}}\!\langle s| \Psi\rangle^{(m,n)}_{A_{1} A_{2}^{'}}|^2.
\label{eq:prob}
\end{equation}

At this stage one may consider various cases by choosing different combinations of $m$ and $n$.
Here we broadly classify all these cases into three categories such as 
\begin{itemize}
\item $m=n$: In this case, the number of input photons is equal to the number
of detected photons in the mode $F_1$. This case is popularly known as photon catalysis~\cite{catalysis1} and 
leads to non-Gaussian states even for $m=n=0$.
\item $m<n$: In this case the number of photons detected in the mode $F_1$ is more than the 
input number of photons in the mode $F_0$. This leads to an overall deduction in the number of photons in the 
original TMSC state leading to a photon subtracted state. Hence the name photon subtraction. The 
resultant state is a non-Gaussian state.
\item $m>n$: In this case, the number of photons detected in the mode $F_1$ is less than 
the number of photons input in the mode $F_0$. This way we can add to the total number of 
photons in the original TMSC state, with the resultant state being non-Gaussian.
\end{itemize}

\begin{figure}
	\includegraphics[scale=1]{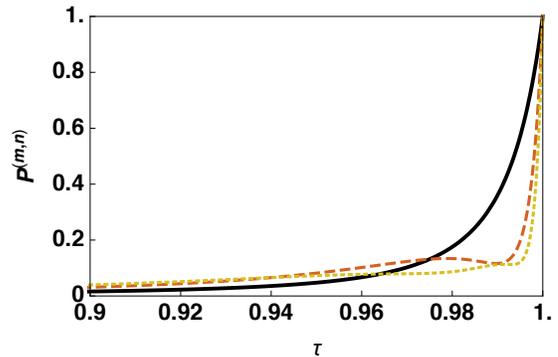}
	\caption{Probability of photon catalysis as a function of BS transmittance $\tau$. The parameters are 
		fixed as: $V_A = 50$ and displacement $d=2$. Various plots correspond to $(0,0)$-CTMSC (Black solid),
		$(1,1)$-CTMSC (Red dashed) and $(2,2)$-CTMSC (Yellow tiny dashed). Plotted parameters are dimensionless.}
	\label{fig:probcatalysis}
\end{figure}

\begin{figure}
	\includegraphics[scale=1]{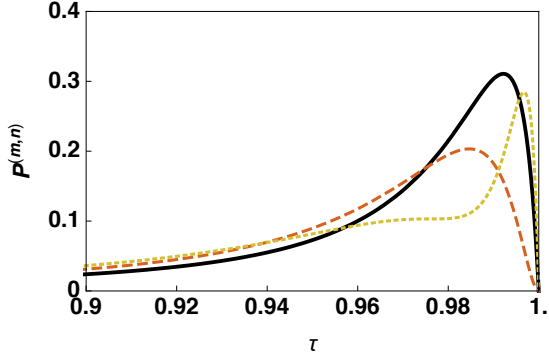}
	\caption{Probability of photon addition as a function of BS transmittance $\tau$. The parameters are 
		fixed as: $V_A = 50$ and displacement $d=2$. Various plots correspond to $(1,0)$-PATMSC (Black solid),
		$(2,0)$-PATMSC (Red dashed) and $(2,1)$-PATMSC (Yellow tiny dashed). Plotted parameters are dimensionless.}
	\label{fig:probaddition}
\end{figure}

\begin{figure}
	\includegraphics[scale=1]{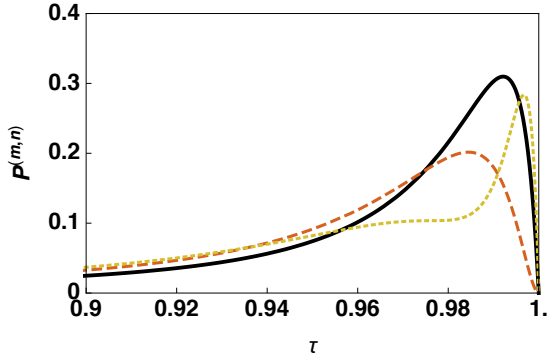}
	\caption{Probability of photon subtraction as a function of BS transmittance $\tau$. The parameters are 
		fixed as: $V_A = 50$ and displacement $d=2$. Various plots correspond to $(0,1)$-PSTMSC (Black solid),
		$(0,2)$-PSTMSC (Red dashed) and $(1,2)$-PSTMSC (Yellow tiny dashed). Plotted parameters are dimensionless.}
	\label{fig:probsubtraction}
\end{figure}

We denote these cases as photon catalyzed TMSC (CTMSC),  photon subtracted TMSC (PSTMSC) and photon added TMSC (PATMSC)  respectively.  
The latter has been studied in depth in Ref.~\cite{chandan_cvmdiqkd}. 
The probability of $n$ photon detections as a function of $\tau$ is plotted in Figs.~\ref{fig:probcatalysis},  \ref{fig:probaddition}
and \ref{fig:probsubtraction} corresponding to CTMSC,  PATMSC and PSTMSC.  
It should be noted that the value of $\tau$ used throughout the paper is optimized to maximize the transmission distance and not photon detection. 

Fred has to publicly announce when the required $(m,n)-$TMSC state has been prepared.  
Thus, it is natural to assume that any and all modes of Fred can be accessed by an eavesdropper, Eve.  
This also allows us to have the device with Fred to be fully uncharacterized such that there may exist information side-channels to Eve.  
Thus, for the remainder of this paper we assume that Fred is an untrusted party and separate from Alice.

The position of Fred plays also plays an important role in non-Gaussian CV MDI QKD. 
Fred can either be placed only on Alice's side (before Charlie),  on Bob's side (after Charlie) or on both sides.  
However,  placing Fred between Bob and Charlie will not offer any advantage as 
the parties will apply classical reverse reconciliation techniques to extract a secure correlated keyrate. 
In this case, Bob will try to align his bits to that of Alice's. It is therefore the case that Alice's source 
prepares the information that is sent to Bob, while Bob's state is only used to guess the bit of Alice.
Therefore, placing Fred between Bob and Charlie will not provide any benefit. Moreover, placing Fred 
at both the locations (between Alice-Charlie and Bob-Charlie) will eventually be extremely detrimental as the probability of detecting $n$ photons simultaneously on modes $A_2$,  $B_2$  is very low.
Consequently,  we have considered the case where Fred lies between Alice and Charlie.

The location of Fred can be further chosen to be either close to Alice,  in between Alice and Charlie or close to Charlie.  
The main purpose of Fred is to increase the entanglement between the modes $A_1$ and $A'_2$ by performing non-Gaussian operations through photon detection on the Gaussian TMSC state. 
Therefore,  if Fred is close to Charlie,  he will be performing photon detections on a state which has already passed through a lossy channel and undergone noise.  
The state that he receives is a mixed state with less entanglement than the initial TMSC state.  
Therefore,  photon detections on this state will result in a final state with lower entanglement
content than the final state generated if he were located close to Alice.  
This,  in turn,  can lead to lesser keyrate or transmission distance.  
Thus in the rest of this paper we assume that Fred is located close to Alice.

We calculate the covariance matrix of $(m,n)$-TMSC in terms of moment generating functions and 
is given as
\begin{equation}
\Sigma_{A_1 A'_2} = \begin{pmatrix}
V_{A}^{x} & 0 & V_{C}^{x} & 0 \\
0 & V_{A}^{p} & 0 & V_{C}^{p} \\
V_{C}^{x} & 0 & V_{B}^{x} & 0 \\
0 & V_{C}^{p} & 0 & V_{B}^{p}
\end{pmatrix},
\label{eq:variance1}
\end{equation}
where $V_i^\xi$, $i\in\lbrace A,B,C\rbrace$ and $\xi\in\lbrace
x, p\rbrace$ is interpreted as variance of $\xi$ quadrature 
for the $i^{th}$ party (see Appendix~\ref{app:cov}). Fred announces when the 
$(m, n)-$TMSC state has been successfully prepared and consequently the mode $A'_2$ is transmitted to Charlie via a quantum channel. 

\textbf{Step 3:} Bob also prepares a TMSC state with variance $V_B = V_A$ and transmits the 
mode $B_2$ to Charlie.

\textbf{Step 4:} Charlie mixes the two modes $A'_2$ and $B_2$ via a BS with output modes
$C$ and $D$. He then performs a homodyne measurement of $x$ and $p$ quadrature on $C$ and $D$
respectively. The outcomes to these measurements are then declared publicly.

\textbf{Step 5:} Based on the publically declared results, Bob displaces his mode $B_1$ to 
$B'_1$ by applying $D(\alpha)$. As a consequence
the modes $A_1$ and $B'_1$ are entangled. 

\textbf{Step 6:} Alice and Bob perform heterodyne measurements on the entangled 
modes $A_1$ and $B'_1$ to get correlated outcomes.

\textbf{Step 7:} Alice and Bob perform information reconciliation and privacy amplification
to obtain a secure key. Here we follow reverse reconciliation~\cite{grosshans_cv_gaussian} (from Bob to Alice) as it is more secure and is known to perform better~\cite{chen_cvmdi_qkd}.

\subsection{Eavesdropping, channel parameters and secure key
	rate}
\label{subsec:eavesdropping}
In this subsection we describe several parameters which will 
be useful in simulating the secure key rates obtained by Alice
and Bob in the presence of an eavesdropper Eve. While most of the 
terminology has been set up in Ref.~\cite{chandan_cvmdiqkd}, we recap it 
here for brevity of the reader.

The CV MDI QKD protocol detailed above comprises of two 
quantum channels between Alice, Bob and Charlie and a single 
classical channel between Alice and Bob. We assume that Eve can 
perform independent one-mode collective 
attacks~\cite{entangling_cloner, entangling_cloner1,entangling_cloner2} on each 
channel and the maximum information that can be obtained is then quantified 
by the Holevo bound $\chi_{BE}$ between Bob and Eve.

We assume that the two channels have transmittance
$T_A$ and $T_B$, given as, 
\begin{equation} 
T_A=10^{-l
	\frac{L_{AC}}{10}} \quad \text{and} \quad T_B=10^{-l
	\frac{L_{BC}}{10}}, 
\label{eq:transmittance} 
\end{equation}
where $l=0.2$dB/Km is the channel loss. Furthermore, we only 
consider the asymmetric case in which $L_{BC}=0$,
implying Bob and Charlie are at the same place. The total
transmission length is then $L=L_{AC}=L_{AB}$ with
$T_B=1$. The symmetric case in which Charlie is midway between 
Alice and Bob has been found to be subpar than 
the asymmetric one in several previous results~\cite{chandan_cvmdiqkd,ma_cvmdiqkd_photon_sub}.

We define a normalized parameter $T$ associated with
channel transmittance in terms of
$T_A$ as
\begin{equation}
T=\frac{T_A g^2}{2},
\label{eq:T}
\end{equation}
where $g$ is the gain of Bob's displacement operation. Total added 
noise in the channel can then be defined as,
\begin{equation}
\chi_{line} = \frac{1-T}{T}+\varepsilon_{th},
\label{eq:chi_line}
\end{equation}
where $\varepsilon_{th}$ is the thermal excess noise in the
equivalent one-way protocol~\cite{ma_cvmdiqkd_photon_sub} which can be written as
\begin{equation}
\varepsilon_{th}=\frac{T_B}{T_A}\left(\varepsilon_B-2\right)+
\varepsilon_A+\frac{2}{T_A},
\label{eq:epsilon_thermal}
\end{equation}
where $\varepsilon_A$ and $\varepsilon_B$ correspond to
thermal excess noise in the respective quantum channels. The gain is then taken as
\begin{equation}
g=\sqrt{\frac{2\left(V_A-1\right)}{T_B\left(V_A+1\right)}},
\label{eq:displacement}
\end{equation}
in order to minimize $\varepsilon_{th}$.

We also assume that Charlie's homodyne detectors are noisy,
with excess noise given as
\begin{equation}
\chi_{homo}=\frac{v_{el}+1-\eta}{\eta},
\label{eq:chi_homo}
\end{equation}
where, $v_{el}$ is the electric noise of the detectors and
$\eta$ is the efficiency. Therefore, the total noise added
because of the 
channel and detectors is
\begin{equation}
\chi_{tot}=\chi_{line}+\frac{2\chi_{homo}}{T_A}.
\label{eq:keyrate}
\end{equation}

The secure key rate obtained by Alice and Bob is given as,
\begin{equation}
K=P^{(m,n)}\left(\beta I_{AB}-\chi_{BE}\right),
\label{eq:secure_keyrate}
\end{equation}
where $P^{(m, n)}$ is the probability to obtain the $(m, n)-$TMSC state given in Eq.~\eqref{eq:prob}, $I_{AB}$ is the mutual information between Alice and
Bob and $\chi_{BE}$ is the Holevo bound between Bob and Eve. The factor 
$P^{(m,n)}$ appears in Eq.~\eqref{eq:secure_keyrate} because the final $(m, n)-$TMSC state is 
obtained probabilistically depending on the detection of $n$ photons. Thus the final resource is 
a fraction of the initial TMSC state. 

The covariance matrix corresponding to the state $\rho_{A_1B'_1}$ which is 
obtained after Step 5. of the
protocol is 
\begin{equation}
\Sigma_{A_1 B'_1} = \begin{pmatrix}
V_{A}^{x} & 0 & \sqrt{T}V_{C}^{x} & 0 \\
0 & V_{A}^{p} & 0 & \sqrt{T}V_{C}^{p} \\
\sqrt{T}V_{C}^{x} & 0 & TV'^x_{B} & 0 \\
0 & \sqrt{T}V_{C}^{p} & 0 & TV'^p_{B}
\end{pmatrix},
\end{equation}
where $V'^\xi_B = V^\xi_{B}+\chi_{tot}I_2$ and $V_B^\xi$ is the 
variance of $\xi\in\lbrace x,p\rbrace$
quadrature for Bob's state..
The mutual information between Alice and Bob, $I_{AB}$ can
then be calculated as,

\begin{equation}
I_{AB}=\frac{1}{2}\log_2\left(\frac{V_{A_M}^x}{V_{A_M|B_M}^x}\right)+
\frac{1}{2}\log_2\left(\frac{V_{A_M}^p}{V_{A_M|B_M}^p}\right),
\end{equation}
such that,
\begin{equation}
V_{A_M}^\xi = \frac{V_A^\xi+1}{2},
\end{equation} 
where $V_{A_M|B_M}^\xi$ is the conditional variance of 
Alice's outcome conditioned on Bob's outcome of his
heterodyne measurement given by,
\begin{equation}
V_{A_M|B_M}^\xi = \frac{V_{A|B}^\xi+1}{2},
\end{equation}
where, 
\begin{equation}
V_{A|B}^\xi = V_A^\xi-V_C^\xi\left(V_B^\xi+I_2\right)^{-1}(V_C^\xi)^T.
\end{equation}

In order to calculate the Holevo bound $\chi_{BE}$, we assume that Eve 
also has access to Fred's mode $F$ and her state is then
given by $\rho_{EF}$. We also assume that 
she can purify $\rho_{A_1B'_1EF}$. The Holevo bound
$\chi_{BE}$ between Bob and Eve can then be calculated as,
\begin{equation}
\begin{aligned}
\chi_{BE}&=S(\rho_{EF})-\int dm_B p(m_B)S(\rho_{EF}^{m_B})\\
&=S(\rho_{A_1 B'_1}) - S(\rho_{A_1}^{m_{B'_1}}),
\label{eq:holevo}
\end{aligned}
\end{equation}
where $S(\rho)$ is the von-Neumann entropy of the state
$\rho$, $m_B$ represents measurement outcomes of Bob with
probability
density $p(m_B)$ and $\rho_{EF}^{m_B}$ is the state of Eve
conditioned on Bob's outcome. The covariance matrices
corresponding to the 
states $\rho_{A_1B'_1}$ and $\rho_{A_1}^{m_{B'_1}}$ are
represented by $\Sigma_{A_1 B'_1}$ and
$\Sigma_{A_1}^{m_{B'_1}}$ respectively. The 
von-Neumann entropy $S(\rho_{A_1 B'_1})$ and
$S(\rho_{A_1}^{m_{B'_1}})$ are functions of symplectic
eigenvalues $\lambda_1$, $\lambda_2$ of $\Sigma_{A_1 B'_1}$
and $\lambda_3$ of $\Sigma_{A_1}^{m_{B'_1}}$
which are given as,
\begin{equation}
S(\rho_{A_1 B'_1}) = G\left[\frac{\lambda_1-1}{2}\right]+
G\left[\frac{\lambda_2-1}{2}\right],
\label{eq:sab}
\end{equation}
and
\begin{equation}
S(\rho_{A_1}^{m_{B'_1}}) = G\left[\frac{\lambda_3-1}{2}\right],
\label{eq:sa}
\end{equation}
with,
\begin{equation}
G(x)=(x+1)\log_2(x+1)-x\log_2x,
\label{eq:thermalstate}
\end{equation}
is the von-Neumann entropy of the thermal state.

\section{Simulation results}
\label{sec:results}

In this section we provide numerical results corresponding to the aforementioned non-gaussian operations
on a TMSC state. For each case we analyse the effects of coherence and non-Gaussianity on keyrate
and transmission distances. 

\subsection{Effect of displacement for a fixed key rate}

In this subsection we analyse the effect of displacement on transmission distances for fixed keyrate
corresponding to CTMSC, PATMSC and PSTMSC.

\begin{figure}
\includegraphics[scale=0.8]{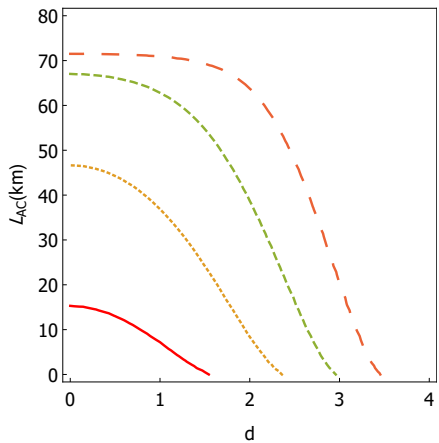}
\caption{Contour plot of displacement $d$ (dimensionless) and transmission distance $L_{AC}$ (km) in the extreme asymmetric case as a 
function of keyrate (bits/pulses) for the case of $(0,0)$-CTMSC. The parameters are fixed as: $V_A = 50$, $\tau = 0.9$, $\epsilon^{th}_A = 0.002 = \epsilon^{th}_B$, $\beta =96\%$.
Various curves correspond to different values of fixed key rate $K=10^{-1}$ (Red solid), $K=10^{-2} $ (Tiny dashed), $K=10^{-3}$ (Dashed) and 
$K=10^{-4}$ (Large dashed). }
\label{fig:dvslcatalysis}
\end{figure}

\begin{figure}
\includegraphics[scale=0.8]{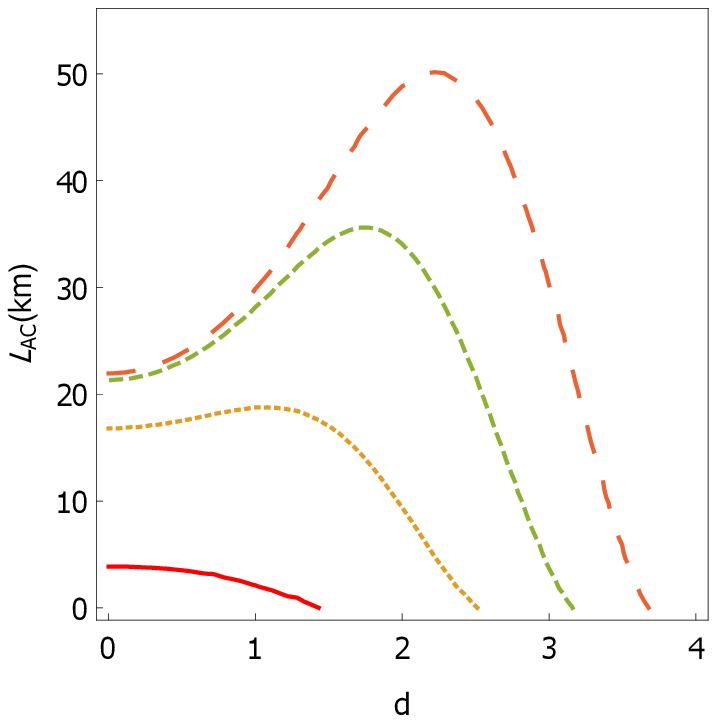}
\caption{Contour plot of displacement $d$ (dimensionless) and transmission distance $L_{AC}$ (km) in the extreme asymmetric case as a 
function of keyrate (bits/pulses) for the case of $(1,0)$-PATMSC. The parameters are fixed as: $V_A = 50$, $\tau = 0.9$, $\epsilon^{th}_A = 0.002 = \epsilon^{th}_B$, $\beta =96\%$.
Various curves correspond to different values of fixed key rate $K=10^{-1}$ (Red solid), $K=10^{-2}$ (Tiny dashed), $K=10^{-3}$ (Dashed) and 
$K=10^{-4}$ (Large dashed). }
\label{fig:dvsladdition}
\end{figure}

\begin{figure}
\includegraphics[scale=0.8]{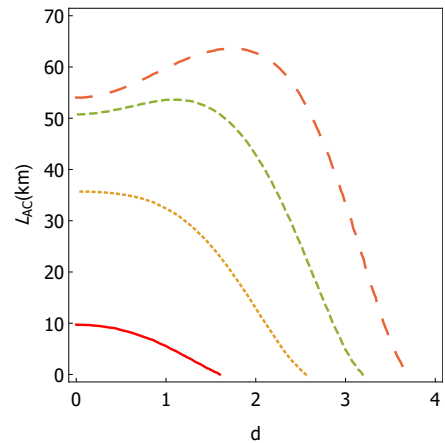}
\caption{Contour plot of displacement $d$ (dimensionless) and transmission distance $L_{AC}$ (km) in 
the extreme asymmetric case as a 
function of keyrate (bits/pulses) for the case of $(0,1)$-PSTMSC. The parameters are 
fixed as: $V_A = 50$, $\tau = 0.9$, $\epsilon^{th}_A = 0.002 = \epsilon^{th}_B$, $\beta =96\%$.
Various curves correspond to different values of fixed 
key rate $K=10^{-1}$ (Red solid), $K=10^{-2}$ (Tiny dashed), $K=10^{-3}$ (Dashed) and 
$K=10^{-4}$ (Large dashed). }
\label{fig:dvslsubtraction}
\end{figure}

As is evident from Fig.~\ref{fig:dvslcatalysis}, in the case of CTMSC, transmission distance decreases monotonically with increased displacement, with a maximum distance of $70$ km achieved for $K=10^{-4}$ bits/pulses. 
Therefore, catalysis on TMSC or TMSV yields equivalent results with no increase in transmission distances and it is therefore preferable to use minimal or no displacement.
On the other hand, photon addition on TMSC state is advantageous than TMSV as is evident  from Fig.~\ref{fig:dvsladdition}. 
It is seen that transmission distances increase drastically with increasing displacement. 
However, we also note that displacement cannot be increased indefinitely as it begins to have detrimental effects on the transmission distance. 
A maximum distance of $50$ km can be achieved with $K=10^{-4}$ for $d\approx 2$.

\begin{figure}
	\centering
	\begin{minipage}{0.25\textwidth}
		\centering
		\includegraphics[scale=0.6]{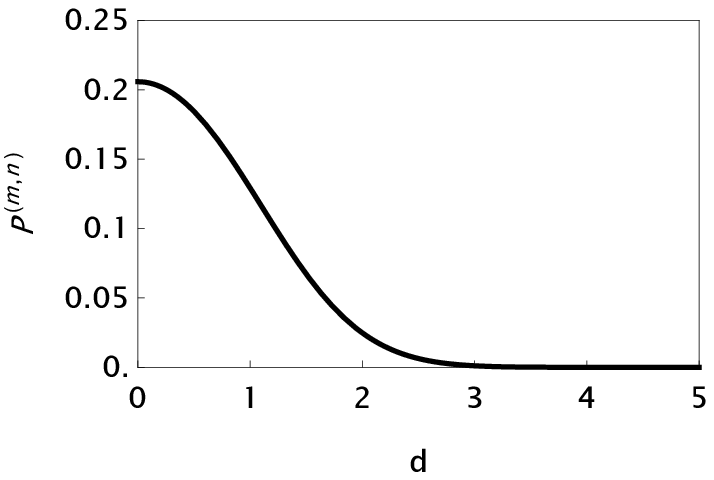}
	\end{minipage}%
	\begin{minipage}{0.25\textwidth}
		\centering
		\includegraphics[scale=0.6]{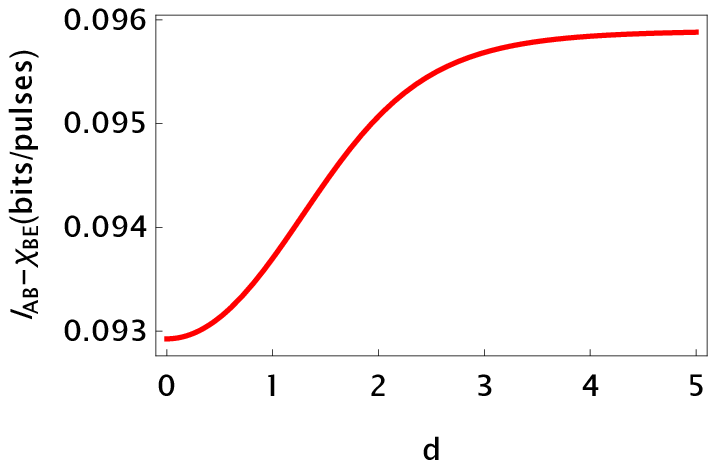}
	\end{minipage}
\caption{Plot of probability $P^{(m,n)}$ and $I_{AB} - \chi_{BE}$ (bits/pulse) with displacement $d$ for 
	the case of $(0,1)-$PSTMSC with parameters 
$V_A = 50, \tau = 0.9, \epsilon^{th}_A = 0.002 = \epsilon^{th}_{B}, \beta = 100\%$ and 
$L_{AC} = L_{AB} = 50$ km. The latter 
plot shows a gradual increase upto a certain maximum value with increasing $d$, while the former
reaches zero just before $d = 3$.}
\label{fig:change_with_d}
\end{figure}

The apparent non-monotonic behaviour of the keyrate with displacement could be understood in terms of the interplay between the success probability ($P^{(m,n)}$) and the difference between mutual information and Holevo information ($I_{AB} - \chi_{BE}$).
Here, we explain for the case of single photon subtracted TMSC - $(0,1)-$PSTMSC.
As can be seen in Fig.~\ref{fig:change_with_d},  for a fixed BS transmittivity $\tau$, with an 
increase in the displacement amplitude ($d$), the probability of photon subtraction drops, while 
simultaneously, the difference between $I_{AB}$ and $\chi_{BE}$ increases.  
This results in an increase in the keyrate ($K=P^{(m,n)}\left(\beta I_{AB}-\chi_{BE}\right)$) upto $d\approx 2$.  
However,  for larger displacement ($d>2$),  while the difference between $I_{AB}$ and $\chi_{BE}$ saturates the success probability falls drastically.  
As a consequence,  the overall keyrate falls beyond the optimal displacement which in our case is $d=2$. 

Photon subtraction on TMSC has been studied extensively in a previous study~\cite{chandan_cvmdiqkd}. 
For the sake of completion we reproduce the same results, albeit using the generalized covariance matrix as derived in this paper. 
From Fig.~\ref{fig:dvslsubtraction}, we conclude that displacement can effectively increase the transmission distances of CV-MDI QKD 
protocols. 

\subsection{Effect of length on keyrate}

In this subsection we analyse the available keyrate with respect to transmission distances in the 
extreme asymmetric case.

\begin{figure}
	\includegraphics[scale=1]{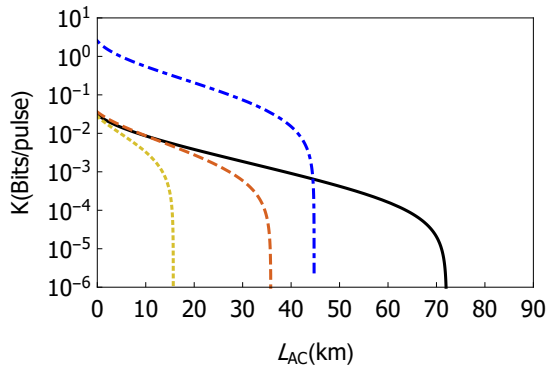}
	\caption{Secret keyrate as a function of $L_{AC}$ in the extreme asymmetric case. The parameters are 
		fixed as: $V_A = 50$, $\tau = 0.9$, $\epsilon^{th}_A = 0.002 = \epsilon^{th}_B$, $\beta =96\%$ and 
		displacement $d=2$. Various plots correspond to TMSV (Blue dash dotted), $(0,0)$-CTMSC (Black solid),
		$(1,1)$-CTMSC (Red dashed) and $(2,2)$-CTMSC (Yellow tiny dashed)}
	\label{fig:lvskcatalysis}
\end{figure}

From Fig.~\ref{fig:lvskcatalysis} we find that the $(0,0)$-CTMSC state offers a dramatic increase in transmission distances as compared to $(1,1)$, $(2,2)$-CTMSC and TMSV state. 
A maximum distance of more than $70$ km can be achieved using the same. 
However, $(1,1)$ and $(2,2)$-CTMSC fare poorly than even the TMSV state.

\begin{figure}
	\includegraphics[scale=1]{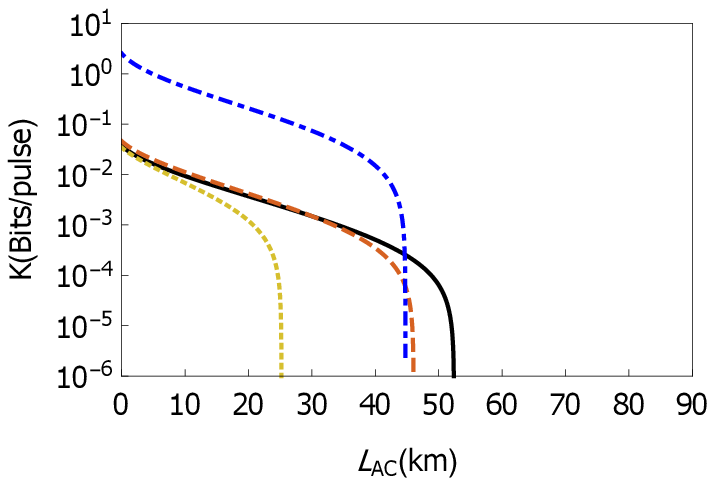}
	\caption{Secret keyrate as a function of $L_{AC}$ in the extreme asymmetric case. The parameters are 
		fixed as: $V_A = 50$, $\tau = 0.9$, $\epsilon^{th}_A = 0.002 = \epsilon^{th}_B$, $\beta =96\%$ and 
		displacement $d=2$. Various plots correspond to TMSV (Blue dash dotted), $(1,0)$-PATMSC (Black solid),
		$(2,0)$-PATMSC (Red dashed) and $(2,1)$-PATMSC (Yellow tiny dashed)}
	\label{fig:lvskaddition}
\end{figure}

As is evident from Fig.~\ref{fig:lvskaddition}, the $(1,0)$-PATMSC state offers better transmission 
distances than $(2,0)$, $(2,1)$-PATMSC and TMSV states. However, the distances are still comparatively 
smaller than what was observed for $(0,0)$-CTMSC state.

\begin{figure}
	\includegraphics[scale=1]{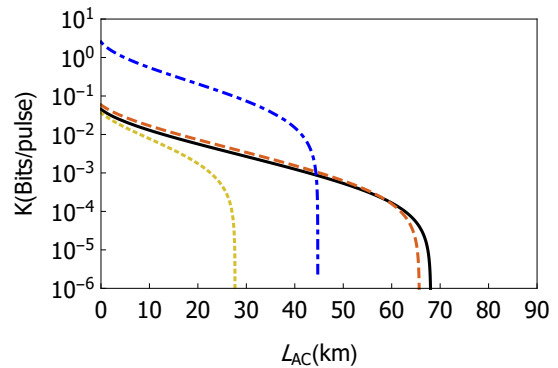}
	\caption{Secret keyrate as a function of $L_{AC}$ in the extreme asymmetric case. The parameters are 
		fixed as: $V_A = 50$, $\tau = 0.9$, $\epsilon^{th}_A = 0.002 = \epsilon^{th}_B$, $\beta =96\%$ and 
		displacement $d=2$. Various plots correspond to TMSV (Blue dash dotted), $(0,1)$-PSTMSC (Black solid),
		$(0,2)$-PSTMSC (Red dashed) and $(1,2)$-PSTMSC (Yellow tiny dashed)}
	\label{fig:lvsksubtraction}
\end{figure}

From Fig.~\ref{fig:lvsksubtraction}, it is clear that $(0,1)$ and $(0,2)$-PSTMSC states offer 
equally good keyrates for large transmission distances than either $(1,2)$-PSTMC or TMSV state. 
It should also be noted that photon subtraction is the only case (considered so far) that offers a 
substantial improvement in transmission distances for single as well as two photon operation.

From the above analysis it is clear that $(0,0)$-CTMSC state offers the highest transmission distance.
However, $(0,1)$ and $(0,2)$-PSTMSC states offer a similar performance. Since, the experimental 
implementation of both is more or less the same, these states should be preferred for CV MDI QKD.

One of the major factors limiting transmission distances (and equivalently the 
secure keyrate) is the noise added to the channel and how it affects each state correspondingly.
More noise will imply smaller transmission distances and vice versa. The channel parameters that 
we have chosen in our plots are achievable in the laboratory while detection ineffeciency with 
Charlie is assumed to zero. In the next subsection we look at the effect of noisy homodyne 
detections with Charlie, which results in added noise in the channel between Alice and Bob.

\subsection{Noisy homodyne detection}

In this subsection we analyse the keyrate under noisy homodyne detectors with Charlie. We observe that under noise
the transmission distances are affected greatly. 

\begin{figure}
\includegraphics[scale=1]{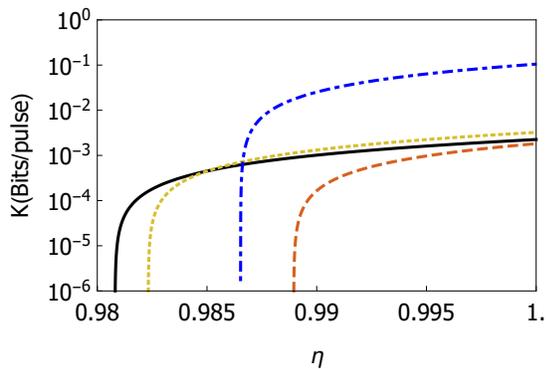}
\caption{Secret keyrate as a function of detection inefficiency $\eta$ (dimensionless) in the extreme asymmetric case. The parameters are 
fixed as: $L_{AC} = 20$ km, $\tau = 0.9$, $\epsilon^{th}_A = 0.002 = \epsilon^{th}_B$, $\beta =96\%$ and 
displacement $d=2$. Various plots correspond to TMSV (Blue dash dotted), $(0,0)$-CTMSC (Black solid),
$(1,0)$-PATMSC (Red dashed) and $(0,1)$-PSTMSC (Yellow tiny dashed).}
\label{fig:etavsknoise}
\end{figure}

From Fig.~\ref{fig:etavsknoise}, we see that $(0,0)$-CTMSC state is the most robust under 
detector noise, while photon addition has the worst response. Photon subtraction is also seen to 
perform adequately as compared to others. It should also be noted that the keyrate for all cases except
TMSV is quite low around approximately $K\approx 10^{-3}$ bits/pulses.

\begin{figure}
\includegraphics[scale=1]{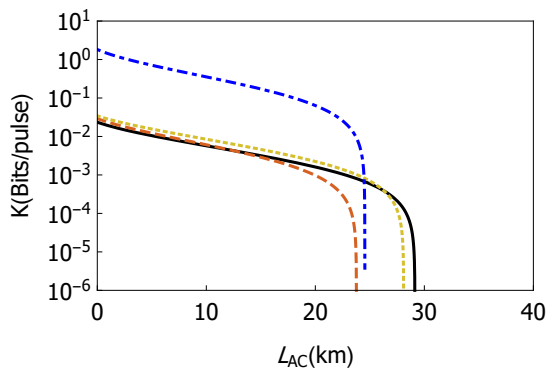}
\caption{Secret keyrate as a function of $L_{AC}$ in the extreme asymmetric case. The parameters are 
fixed as: $\eta = 0.995$, $\tau = 0.9$, $\epsilon^{th}_A = 0.002 = \epsilon^{th}_B$, $\beta =96\%$ and 
displacement $d=2$. Various plots correspond to TMSV (Blue dash dotted), $(0,0)$-CTMSC (Black solid),
$(1,0)$-PATMSC (Red dashed) and $(0,1)$-PSTMSC (Yellow tiny dashed).}
\label{fig:lvskbestnoise}
\end{figure}

The total transmission distance is also seen to suffer under detector noise in Fig.~\ref{fig:lvskbestnoise}. 
A maximum distance of 
approximately $29$ km can be achieved by using $(0,0)$-CTMSC, while $(0,1)$-PSTMSC is not far
behind. It is again observed that the photon added state performs even worst than the TMSV state.

\section{Conclusion}
\label{sec:conclusion}

In this paper we derived a generalized covariance matrix for non-Gaussian states comprising of CTMSC, PATMSC and PSTMSC. 
The number of photons to be catalysed, added or subtracted as well as squeezing and displacement are taken as parameters to this covariance matrix. 
Using the generalized covariance matrix we analyse performance of the aforementioned non-Gaussian states in CV MDI QKD. 
We find that $(0,0)$-CTMSC state offers the best possible choice of state as it affords a longer transmission distance and is 
robust against white noise. 
However, $(0,1)$-PSTMSC is also equivalently good. 
We found that PATMSC states are not an optimal choice in CV MDI QKD, but are still better than standard Gaussian states in some cases. 

We also reinforce the fact that coherence is a useful phenomena in increasing the total transmission distances in CV MDI QKD protocols. 
While the effect of displacement has been studied extensively in Ref.~\cite{chandan_cvmdiqkd} for the case of photon subtraction, we 
further generalize it to photon addition and catalysis too. 
In comparison to the earlier studies on quantum catalysis on 
TMSV~\cite{cvcatalysis,ye_cvmdi_qkd_catalysis,wang_cvmdi_qkd_catalysis,hu_catalysis_entanglement}, here we show that additional coherence boosts the performance further. 
However, it must be noted that all these non-Gaussian operations are probabilistic and subject to the finesse of the experimental setup.

The efficacy of photon catalysis operation with displacement could further be cherished under realistic conditions such as imperfect state preparation~\cite{wang_cvmdi_qkd_source} that is abundant in any practical setup.
Moreover, in recent years, there have a been several new proposals for tweaking the modulation to further optimize the keyrate-vs-transmission distance, such as discrete modulation~\cite{ma_cvmdi_qkd_modulation,ye_cvmdi_qkd_catalysis2}, simultaneous classical communication~\cite{wu_cvmdi_qkd_detector}, phase-modulation~\cite{liao_cvmdi_qkd_plugplay} {\em etc.}. 
These render, to the current work immediate relevance and immense interest in present context as well as in other areas of 
continuous variable quantum information processing~\cite{flamini_review}.

\acknowledgements
Both the authors wish to thank Prof. Arvind for helpful discussions regarding non-Gaussian operations
on coherent states.

\section*{Appendix}

\appendix
	
	\section{Wigner Distribution of ($m,n$)-TMSC}
		\begin{figure}[h]
		\centering
		\includegraphics[scale=0.7]{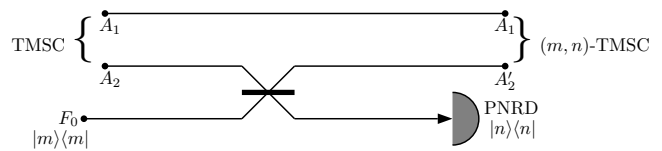}
		\caption{Schematic diagram of generation of $(m,n)$-TMSC state.}
		\label{fig_machine}
	\end{figure}		

	In Fig. \ref{fig_machine}, we portray the generation of $(m,n)$-TMSC pictorially.
	Now we present stepwise calculation of Wigner function for $(m,n)$-TMSC and the corresponding probability in shot noise unit (SNU).
		
\subsection{Wigner Distribution for TMSC}
\label{appwigner:tmsc}
		Let's consider a two mode coherent state, $\rho_{A_{1}A_{2}}^{C}=\ket{d,d}\bra{d,d}$, represented by the Wigner distribution
		\begin{equation}\label{eq:wignercovariance}
		W_{A_{1}A_{2}}^{C}(\xi) = \frac{\exp[-(1/2)(\xi-\overline{\xi})^TV^{-1}(\xi-\overline{\xi})]}{(2 \pi)^2 \sqrt{\text{det}V}},
		\end{equation}
		where $\xi = (x_1, p_1,x_2, p_2)^T$ is the column vector with mode quadratures as its components, $\overline{\xi} = (d,0,d,0)^T$ denotes the corresponding displacement vector and $V = \mathds{1}_2\bigoplus\mathds{1}_2$  is the covariance matrix corresponding to the vacuum state.
		Here, $\mathds{1}_2$ denotes the $2\times 2$ identity matrix. 
		Thus Eq.~(\ref{eq:wignercovariance}) can be explicitly written as
		\begin{equation}
		W^{C}_{A_1 A_2}(\xi) = \frac{1}{4 \pi^2}  e^{ -\frac{1}{2} \left((x_1-d)^2 + p_1^2 + (x_2-d)^2 + p_2^2 \right)}
		\end{equation}
		Now the two mode squeezing transformation is given by
		\begin{equation}
		S_{12}(r) = \begin{pmatrix}
		\cosh r\, \mathds{1}_{2 } & \sinh r\, \mathbb{Z}_{2 } \\
		\sinh r\, \mathbb{Z}_{2 }& \cosh r\, \mathds{1}_{2 }
		\end{pmatrix},
		\end{equation}
		where $\mathbb{Z}_2 = \begin{pmatrix} 
		1&0\\
		0&-1
		\end{pmatrix}$.
		Under the transformation $S_{12}(r)$, Wigner distribution changes as $S_{12}(r): W^{C}_{A_1 A_2}(\xi)\rightarrow  W_{A_1 A_2}(\xi) = W^{C}_{A_1 A_2}(S_{12}^{-1}(r)\xi)$ , i.e.,
		\begin{widetext}
			\begin{equation}
			\begin{aligned}
			W_{A_1 A_2}(\xi) = \frac{1}{4 \pi^2}  \exp \bigg[-\frac{1}{2} &((x_1 \cosh \, r-x_2 \sinh \, r-d)^2
			+(p_1 \cosh \, r+p_2 \sinh \, r)^2\\
			&
			+(x_2 \cosh \, r-x_1 \sinh \, r-d)^2
			+(p_1 \sinh \, r+p_2\cosh \, r)^2
			)\bigg]
			\end{aligned}
			\end{equation}
		\end{widetext}
		
		\subsection{Wigner Distribution for $(m,n)$-TMSC}\label{app:wignerbeam}
		
		Fred mixes the ancilla mode $F_0$ in number state $\ket m$ with mode $A_2$ of TMSC using a beam splitter of transmittivty $\tau$ - represented by the transformation matrix
		\begin{equation}
		B(\tau) = \begin{pmatrix}
		\sqrt{\tau}\mathds{1}_{2 } & \sqrt{1-\tau}\mathds{1}_{2 } \\
		-\sqrt{1-\tau}\mathds{1}_{2 }& \sqrt{\tau}\mathds{1}_{2 }
		\end{pmatrix}.
		\end{equation}
		
		This introduces the transformation 
		\begin{equation}
		S_{\rm{BS}}=\mathds{1}_{2}\bigoplus B(\tau)
		\end{equation} 
		on the three mode quadrature vector $\tilde{\xi} = (x_1, p_1,x_2, p_2,x_3,p_3)^T$ for the input state described by the Wigner distribution $W_{A_1 A_2 F_0}(\tilde{\xi}) = W_{A_1 A_2}(\xi) \otimes W_{F_0}^{\ket m}(\xi_3) = W_{A_1 A_2}(\xi_1,\xi_2) \otimes W_{F_0}^{\ket m}(\xi_3)$, where $\xi_i = (x_i,p_i)^{T} (i=1,2,3)$ and $W^{\ket m}(\xi_3)$ is the Wigner distribution for number state $\ket m$ given as 
		\begin{widetext}
			\begin{equation}
			\label{eq:outputfock}
			W^{\ket m}(x_3,p_3)= \frac{(-1)^m}{2 \pi} e^{-\frac{x_3^2 + p_3^2}{2}}\left.  \partial_{s}^{n}\partial_t^n(e^{st+s(x_3+ip_3)-t(x_3-ip_3)})\right|_{s=t=0} .
			\end{equation}
		\end{widetext}
		
		Consequently, the BS input three mode Wigner distribution changes as $S_{\rm{BS}}: W_{A_1 A_2 F_0}(\tilde{\xi}) \rightarrow W_{A_1 A_2^{'} F_1}(\tilde{\xi}) = W_{A_1 A_2 F_0}(S_{\rm{BS}}^{-1} \tilde{\xi}) =  W_{A_1 A_2}(\xi_1,\xi_2^{'}) W_{F_0}^{\ket m}(\xi_{3}^{'})$.
		
		After a successful detection of $n$-photons i.e. when $\Pi = |n\rangle \langle n|$ clicks, the unnormalized Wigner distribution for $(m,n)$-TMSC becomes
		\begin{equation}
		\begin{aligned}
		W^{(m,n)}_{A_1 A_2'}(\xi_1,\xi_2) &= 4 \pi \int dx_3 dp_3 \, W_{A_1 A_2}(\xi_1, \xi_2^{'})
		W_{F_0}^{\ket m}(\xi_3^{'}) \,
		\\
		& \quad \quad \quad \quad W^{\ket n}_{F_0}(\xi_3).
		\end{aligned}
		\end{equation}
		
		
		As we shall see, we do not need to explicitly calculate the Wigner distribution for $(m,n)$-TMSC in our probability and covariance matrix calculation. 
		
		\section{Calculation of Probability of $(m,n)$-TMSC}\label{app:prob}
		
		The probability of $n$-photon detection is obtained by integrating $W^{(m,n)}_{A_1 A_2'}(\xi_1,\xi_2)$ as
		\begin{equation}\label{eq:2ndprobability}
		\begin{aligned}
		P^{(m,n)} &= \int \xi_1 \xi_2 \, W^{(m,n)}_{A_1 A_2'}(\xi_1,\xi_2)\\
		&= 4 \pi\int d^6 \tilde{\xi} \,W_{A_1 A_2}(\xi_1,\xi_2^{'}) W^{\ket m}_{F_0}(\xi_3^{'}) W^{\ket n}_{F_0}(\xi_3),
		\end{aligned}
		\end{equation}
		
		Now, using the generating function of Laguerre polynomial
		\begin{equation}
		\left.  \partial_{s}^{k}\partial_t^k(e^{st+s(q+ip)-t(q-ip)})\right|_{s=t=0} = k!L_k(q^2+p^2),
		\end{equation}
		%
		we get the probability of $n$-photon detection as
		\begin{equation}
		\begin{aligned}
		P^{(m,n)} &= \frac{(-1)^{m+n}}{4 \pi^3}\frac{1}{m!n!}e^{-d^2} \partial_{u}^{m}\partial_v^m\partial_{s}^{n}\partial_t^n e^{st+uv} 
		\times
		\\
		&~~~\int d^6 \tilde{\xi} \exp \big(-\tilde{\xi}^T M \tilde{\xi}+N^T \tilde{\xi} \big)\Big|_{u=v=s=t=0},
		\label{eq:multidimensional}
		\end{aligned}
		\end{equation}
		with
		\begin{align}
		M &= \begin{pmatrix}
		m_1 \mathds{1}_2 & m_4 \mathbb{Z}_2 & m_5 \mathbb{Z}_2  \\
		m_4 \mathbb{Z}_2 & m_2 \mathds{1}_2 & m_6 \mathds{1}_2 \\
		m_5 \mathbb{Z}_2& m_6 \mathds{1}_2 & m_3 \mathds{1}_2
		\end{pmatrix} ~~\&
		\\
		\nonumber
		N &= \begin{pmatrix}
		-d n_1  \\
		0\\
		( u- v)\sqrt{1-\tau } -d n_1 \sqrt{\tau}\\
		i (u+v)\sqrt{1-\tau } \\
		s-t+(u-v)\sqrt{\tau}+d n_1 \sqrt{1-\tau}\\
		i(s+t)+i( u+ v)\sqrt{\tau }
		\end{pmatrix},
		\end{align}
		
		where  $m_1= -(1+2 \alpha^2)/2$, $m_2= -(1+2 \alpha^2 \tau)/2$, $m_3= -(1+ \alpha^2(1-\tau))$, $m_4= \alpha\sqrt{(1+\alpha^2)\tau}$, $m_5=- \alpha\sqrt{(1+\alpha^2)(1-\tau)}$, $m_6 = \alpha^2 \sqrt{\tau(1-\tau)}$ and
		$n_1 =\alpha-\sqrt{1+\alpha^2} $ and $\alpha = \sinh \,r $. 
		This form facilitates the use of multidimensional Gaussian integral formula 
		\begin{widetext}
			\begin{equation}
			\int_{\mathbb{R}^n} \exp\big(-X^T M X+N^T X\big) dX = \sqrt{\frac{\pi^n}{\text{det}M}}\exp\bigg( \frac{N^T M^{-1}N}{4}\bigg).
			\end{equation}
		\end{widetext}
		
		Consequently the expression of probability reduces to
		\begin{widetext}
			\begin{equation}
			\begin{aligned}
			P^{(m,n)} &= \frac{(-1)^{m+n}}{m!n!}\frac{1}{1+\alpha^2(1-\tau)}e^{-i_1} \partial_{u}^{m}\partial_v^m\partial_{s}^{n}\partial_t^n e^{-a_1 s t+b_1 s+c_1 t-d_1 u v+e_1 u+f_1 v+g_1 t u+h_1 s v}\bigg|_{u=v=s=t=0}
			\\
			&= \frac{(-1)^{m+n}}{m!n!}\frac{1}{1+\alpha^2(1-\tau)}e^{-i_1} \partial_{u}^{m}\partial_v^m\partial_{s}^{n}\partial_t^n \sum _{l=0}^{\infty } \frac{\left(g_1 t u\right){}^l}{l!} \sum _{k=0}^{\infty } \frac{\left(h_1 s v\right){}^k}{k!} e^{-a_1 s t+b_1 s+c_1 t}e^{-d_1 u v+e_1 u+f_1 v}\bigg|_{u=v=s=t=0}
			\\
			&= \frac{(-1)^{m+n}}{m!n!}\frac{1}{1+\alpha^2(1-\tau)}e^{-i_1} \partial_{u}^{m}\partial_v^m\partial_{s}^{n}\partial_t^n \sum _{l=0}^{\infty }\sum _{k=0}^{\infty } \frac{g_1^l}{l!}  \frac{h_1^k}{k!}\partial_{c_1}^{l}\partial_{b_1}^k\partial_{e_1}^{l}\partial_{f_1}^k e^{-a_1 s t+b_1 s+c_1 t}e^{-d_1 u v+e_1 u+f_1 v}\bigg|_{u=v=s=t=0},
			\end{aligned}
			\label{eq:pcatalysis}
			\end{equation}
			where
			\begin{equation}
			\begin{aligned}
			a_1 &= \frac{\alpha^2}{1+\alpha^2}d_1=\frac{\alpha^2(1-\tau)}{1+\alpha^2(1-\tau)},\\
			c_1&=-b_1=\frac{a_1(\alpha+\sqrt{1+\alpha^2})\sqrt{1-\tau}}{2(1+\alpha^2(1-\tau))},\\
			e_1&=-f_1=\frac{a_1(\alpha+\sqrt{1+\alpha^2})\sqrt{\tau(1-\tau)}}{2(1+\alpha^2(1-\tau))},\\
			\end{aligned}
			\qquad
			\begin{aligned}
			g_1&=h_1=\frac{-\sqrt{\tau}}{1+\alpha^2(1-\tau)},\\
			i_1&=\frac{d^2(1+2 \alpha (\alpha+\sqrt{1+\alpha^2}))(1-\tau)}{4(1+\alpha^2(1-\tau))}.\\
			\end{aligned}
			\end{equation}
		\end{widetext}
		
		Now we recall the following identities for two-variable Hermite polynomial
		\begin{widetext}
			\begin{align}
			\nonumber
			H_{m,n}(x, y)&=\partial_{s}^{m}\partial_t^n \exp (-s t+s x+t y)\big|_{s=t=0} \, \sum _{j=0}^{\min (m,n)} \frac{(-1)^j m! n! x^{m-j} y^{n-j}}{j! (m-j)! (n-j)!} ~~\&
			\\
			\partial_{x}^{k}\partial_y^l   H_{m,n}(x, y)&=\frac{m! n!}{ (m-k)!(n-l)!} H_{m-k,n-l}(x, y)
			\end{align}
		\end{widetext}
		
		These identities reduce Eq. (\ref{eq:pcatalysis})
		\begin{widetext}
			\begin{equation}
			\begin{aligned}
			P^{(m,n)}&= \frac{(-1)^{m+n}}{m!n!}\frac{1}{1+\alpha^2(1-\tau)}e^{-i_1} \sum _{l=0}^{\infty }\sum _{k=0}^{\infty } \frac{g_1^l}{l!}  \frac{h_1^k}{k!}\partial_{c_1}^{l}\partial_{b_1}^k\partial_{e_1}^{l}\partial_{f_1}^k a_1^mH_{m,m}\bigg[\frac{b_1}{\sqrt{a_1}},\frac{c_1}{\sqrt{a_1}}\bigg] d_1^n H_{n,n}\bigg[\frac{e_1}{\sqrt{d_1}},\frac{f_1}{\sqrt{d_1}}\bigg]
			\\
			&=\frac{(-1)^{m+n}}{m!n!}\frac{1}{1+\alpha^2(1-\tau)}e^{-i_1} \sum _{l=0}^{\min (m,n) }\sum _{k=0}^{\min (m,n) } \frac{g_1^l}{l!}  \frac{h_1^k}{k!} a_1^m d_1^n \frac{1}{\sqrt{a_1}^{k+l}} \frac{1}{\sqrt{d_1}^{k+l}}
			\frac{m! m! n! n!}{(m-k)! (n-k)! (m-l)! (n-l)!}
			\\
			&\quad \quad H_{m-k,m-l}\bigg[\frac{b_1}{\sqrt{a_1}},\frac{c_1}{\sqrt{a_1}}\bigg]  H_{n-l,n-k}\bigg[\frac{e_1}{\sqrt{d_1}},\frac{f_1}{\sqrt{d_1}}\bigg]
			\end{aligned}
			\end{equation}
		\end{widetext}
		
		\section{Calculation of covariance matrix }\label{app:cov}
		
		Here we provide a general expression for the moment generating function defined as $\mathscr{F}_{M} = \frac{1}{2}\langle \{\hat{x_1}^{r_1} \hat{p_1}^{s_1} \hat{x_2}^{r_2} \hat{p_2}^{s_2}\}_{\text{sym}}\rangle$. 
		Any particular moment, i.e., the elements of the covariance matrix could be easily obtained from this generating function as special cases, e.g., $\frac{1}{2} \langle \{\hat{x_1} \hat{p_1} \}_{\text{sym}}\rangle = \frac{1}{2} \langle \{\hat{x_1},\hat{p_1} \}\rangle = \lim_{\substack{r_{1}\rightarrow 1, s_{1}\rightarrow 1 \\ r_{2}\rightarrow 0, s_{2}\rightarrow 0}} \mathscr{F}_{M}$, where "$\{ , \}$" denotes the anti-commutator.
		In terms of this normalized Wigner distribution of $(m,n)$-TMSC, $\tilde{W}^{(m,n)}_{A_1 A_2'}(\xi_1,\xi_2) = \frac{1}{P^{(m,n)}}  W^{(m,n)}_{A_1 A_2'}(\xi_1,\xi_2)$, the moment generating function $\mathscr{F}_{M}$ could be easily evaluated by using parametric differentiation techniques as
		\begin{widetext}
			\begin{equation}\label{eq:covfinal}
			\begin{aligned}
			\mathscr{F}_{M} =& \int d^4 \xi x_1^{r_1} p_1^{s_1} x_2^{r_2} p_2^{s_2} \tilde{W}^{(m,n)}_{A_1 A_2'}(\xi_1,\xi_2)
			\\
			&=\frac{1}{P^{(m,n)}} \frac{(-1)^{m+n}}{m!n!}\frac{1}{1+\alpha^2(1-\tau)}\sum _{k,l=0}^{\min (m,n) } \frac{g_1^l}{l!}  \frac{h_1^k}{k!} a_1^{m-\frac{k+l}{2}} d_1^{n-\frac{k+l}{2}}
			\frac{m! m! n! n!}{(m-k)! (n-k)! (m-l)! (n-l)!}
			\times
			\\
			&\partial_{u_1}^{r_1}\partial_{v_1}^{s_1}\partial_{u_2}^{r_2}\partial_{v_2}^{s_2}
			e^{g_2 u_1+ h_2 u_2 +i_2(u_1^2+v_1^2+u_2^2+v_2^2)-j_2(u_1 u_2-v_1 v_2)+k_2}
			\times
			\\
			&H_{m-k,m-l}\bigg[\frac{-a_2 (u_1-i  v_1)-b_2 (u_2+i  v_2)-c_2}{\sqrt{a_1}},\frac{a_2 (u_1+i v_1)+b_2 (u_2-i  v_2)+c_2}{\sqrt{a_1}}\bigg] 
			\times
			\\
			&  H_{n-l,n-k}\bigg[\frac{-d_2 (u_1+i  v_1)-e_2 (u_2- i  v_2)-f_2}{\sqrt{d_1}},\frac{d_2( u_1-i  v_1)+e_2 (u_2+ i v_2)+f_2}{\sqrt{d_1}}\bigg]\Bigg|_{u_1=v_1=u_2=v_2=0}
			\end{aligned}
			\end{equation}
			where,
			\begin{equation}
			\begin{aligned}
			a_2=&\frac{d_2}{\sqrt{\tau}}=\frac{\alpha  \sqrt{(1+\alpha ^2) (1-\tau) }}{1+\alpha ^2 (1-\tau )}, 
			\quad b_2=\frac{\alpha^2\sqrt{\tau (1-\tau)}}{1+\alpha ^2 (1-\tau )},
			\quad c_2=\frac{f_2}{\sqrt{\tau}}=\frac{d (\alpha+\sqrt{1+\alpha^2})\sqrt{1-\tau}}{2(1+\alpha ^2 (1-\tau ))},
			\\
			e_2=&\frac{(1+\alpha^2)\sqrt{1-\tau}}{1+\alpha ^2 (1-\tau )},
			\quad g_2=-\frac{d (\sqrt{1+\alpha^2} +\alpha \tau)}{1+\alpha ^2 (1-\tau )},
			\quad h_2=-\frac{d (\alpha+\sqrt{1+\alpha^2})\sqrt{ \tau}}{1+\alpha ^2 (1-\tau )},\\
			i_2=&-\frac{1+\alpha^2(1+\tau)}{2(1+\alpha ^2 (1-\tau ))} ,
			\quad j_2=\frac{2 \alpha\sqrt{(1+\alpha^2)\tau}}{1+\alpha ^2 (1-\tau )},
			\quad k_2=\frac{d^2(1+2 \alpha(\alpha+\sqrt{1+\alpha^2}))(1-\tau)}{4(1+\alpha ^2 (1-\tau ))}.
			\end{aligned}
			\end{equation}
		\end{widetext}
		
		By suitably choosing values of $r_1$, $s_1$, $r_2$, $s_2$ in Eq.~(\ref{eq:covfinal}), one can calculate all the elements of the covariance matrix that takes the following form:
		\begin{equation}
		\Sigma= (V_{ij})
		\equiv \begin{pmatrix}
		V_{A}^{x} & 0 & V_{C}^{x} & 0 \\
		0 & V_{A}^{p} & 0 & V_{C}^{p} \\
		V_{C}^{x} & 0 & V_{B}^{x} & 0 \\
		0 & V_{C}^{p} & 0 & V_{B}^{p}
		\end{pmatrix}.
		\end{equation}


%

\end{document}